\documentclass[fleqn,usenatbib]{mnras}

\usepackage{newtxtext,newtxmath}

\usepackage[T1]{fontenc}
\usepackage{adjustbox}

\DeclareRobustCommand{\VAN}[3]{#2}
\let\VANthebibliography\thebibliography
\def\thebibliography{\DeclareRobustCommand{\VAN}[3]{##3}\VANthebibliography}

\usepackage{graphicx}	
\usepackage{amsmath}	
\usepackage{array} 
\usepackage{float}

\newcommand{\muas}{\ensuremath{\mu\text{as}}}

\title[Black Hole Image Generation With GANs]{Generating Images of the M87* Black Hole Using GANs}

\author[A. Mohan et al.]{Arya Mohan,$^{1}$\thanks{E-mail: arya.mohan@outlook.com}
Pavlos Protopapas,$^{2}$
Keerthi Kunnumkai,$^{3}$
Cecilia Garraffo,$^{4}$
\newauthor
Lindy Blackburn,$^{5,6}$
Koushik Chatterjee,$^{5,6}$
Sheperd S. Doeleman,$^{5,6}$
Razieh Emami,$^{6}$
\newauthor
Christian M. Fromm,$^{7,8,9}$
Yosuke Mizuno,$^{10,11,12}$
and Angelo Ricarte$^{5,6}$
\\
$^{1}$Univ.AI, Singapore\\
$^{2}$John A. Paulson School of Engineering and Applied Science, Harvard University, Cambridge, MA, 02138, USA\\
$^{3}$Department of Physics, Carnegie Mellon University, Pittsburgh, PA, 15213, USA\\
$^{4}$AstroAI at the Center for Astrophysics | Harvard \& Smithsonian, 60 Garden St. Cambridge, MA, 02138, USA\\
$^{5}$Black Hole Initiative at Harvard University, 20 Garden St., Cambridge MA, 02138\\
$^{6}$Center for Astrophysics | Harvard \& Smithsonian, 60 Garden St. Cambridge MA, 02138\\
$^{7}$Institut f\"ur Theoretische Physik und Astrophysik, Universit\"at W\"urzburg, Emil-Fischer-Strasse 31, 97074
W\"urzburg, Germany\\
$^{8}$Institut f\"ur Theoretische Physik, Goethe Universit\"at, Max-von-Laue-Str. 1, D-60438 Frankfurt, Germany\\
$^{9}$Max-Planck-Institut f\"ur Radioastronomie, Auf dem H\"ugel 69, D-53121 Bonn, Germany\\
$^{10}$Tsung-Dao Lee Institute, Shanghai Jiao-Tong University, Shanghai, 520 Shengrong Road, 201210, People's Republic of China\\
$^{11}$School of Physics \& Astronomy, Shanghai Jiao-Tong University, Shanghai, 800 Dongchuan Road, 200240, People's Republic of China\\
$^{12}$Institut f\"{u}r Theoretische Physik, Goethe Universit\"{a}t, Max-von-Laue-Str. 1, 60438 Frankfurt am Main, Germany}

\date{Accepted XXX. Received YYY; in original form ZZZ}

\pubyear{2023}

\begin{document}
\label{firstpage}
\pagerange{\pageref{firstpage}--\pageref{lastpage}}
\maketitle

\begin{abstract}
In this paper, we introduce a novel data augmentation methodology based on Conditional Progressive Generative Adversarial Networks (CPGAN) to generate diverse black hole (BH) images, accounting for variations in spin and electron temperature prescriptions. These generated images are valuable resources for training deep learning algorithms to accurately estimate black hole parameters from observational data. Our model can generate BH images for any spin value within the range of [-1, 1], given an electron temperature distribution. To validate the effectiveness of our approach, we employ a convolutional neural network to predict the BH spin using both the GRMHD images and the images generated by our proposed model. Our results demonstrate a significant performance improvement when training is conducted with the augmented dataset while testing is performed using GRMHD simulated data, as indicated by the high $R^2$ score. Consequently, we propose that GANs can be employed as cost effective models for black hole image generation and reliably augment training datasets for other parameterization algorithms.
\end{abstract}




\section{Introduction}

\begin{figure}
    \includegraphics[width=\columnwidth]{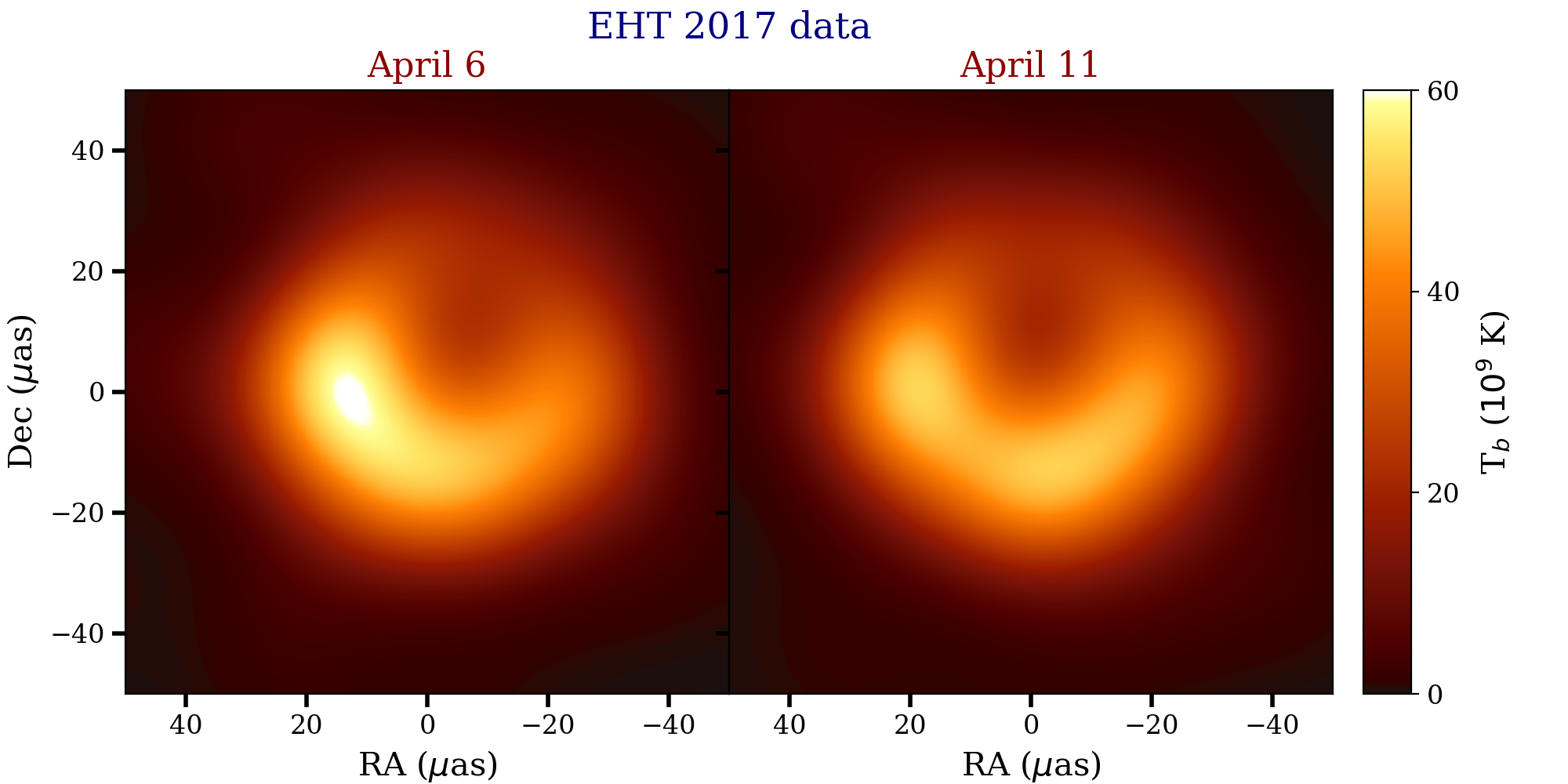}
    \caption{The Event Horizon Telescope (EHT) captured these image using a technique called very-long-baseline interferometry (VLBI). Eight radio telescopes were combined, each operating at millimeter and submillimeter wavelengths, to create a virtual telescope as big as the Earth. With an angular resolution of 20 $\mu$as when operating at 1.3mm, EHT \citep{eht_m87_p2} obtained an image of the M87* at the center of the Messier 87 galaxy.}
    \label{fig:BH}
\end{figure}

One of the key predictions of Einstein’s theory of general relativity (GR) is the existence of black holes (BHs). The synchrotron emission by the hot, orbiting electrons near the BH’s event horizon provides a novel way to directly observe them \citep{v1921lorentz}, \citep{bardeen1973}, which is manifested through a bright ring of emission, singling an interior dark BH ‘shadow’ \citep{luminet1979image,Falcke_2000_shadow}. The shadow diameter for a BH of mass $M$ is roughly 10 $G M/ c^2$, depending on the BH spin as well as the observer inclination. For nearby low-luminosity active galactic nuclei (AGN) sources, initial theory, simulations and observations indicated that such a shadow could be observed at millimeter wavelengths where hot plasma surrounding the black hole becomes transparent \citep{noble2007grmhd,Doeleman_2008,Broderick_2009,Moscibrodzka_2012,Doeleman_2012,yuan2014hot}.  At a distance of $D \simeq 16.8 \ \mathrm{Mpc}$, with a mass of $M \simeq 6.5 \times 10^9 M_{\odot} $, the supermassive black hole at the center of galaxy M87 has a shadow diameter that can be resolved by millimeter wavelength very-long-baseline interferometry (VLBI). The Event Horizon Telescope Collaboration \citep[EHTC][]{eht_m87_p1,eht_m87_p2,eht_m87_p3,eht_m87_p4,eht_m87_p5,eht_m87_p6} captured the first images of M87's shadow, revealing an asymmetric ring-like structure with angular diameter of 42 $\pm$ 3 $\muas$, consistent with predictions of GR (Fig.~\ref{fig:BH}).  Further, in 2022, the EHTC  published the first image of the shadow surrounding Sagittarius A* (SgrA*), the black hole at the center of our own Milky Way galaxy \citep{eht_sgra_p1}.

\vspace{\baselineskip}
\noindent
To model the appearance of M87*, EHT ran general relativistic hydrodynamic (GRMHD) simulations that described the complex dynamics of the gas and plasma surrounding the black hole. The GRMHD simulations are parameterized by the black hole spin ($a_*$), with the simulation library consisting of 5 values of this spin parameter. Next, the EHT performed general relativistic ray tracing (GRRT) based on these GRMHD simulations to create a library of synthetic black hole images to compare directly with the observed image of M87* \citep{eht_m87_p5}. Further, to calculate the synchrotron emission from the simulations, one needs the electron temperature ($T_e$). Since GRMHD simulations only evolve a single temperature fluid, we use the $R-\beta$ prescription \citep{moscibrodzka2016} to define a relationship between the electron temperature and the GRMHD fluid properties based on the local gas plasma-$\beta$, which is the ratio of the gas and magnetic pressures: 

\begin{equation}
    T_{\rm e} = \frac{2 m_{\rm p}U_{\rm gas}}{3k_{\rm B}\rho (2 + R)}
\end{equation}
\\
where
\\
 \begin{equation}
     R = \frac{1+R_\mathrm{high} \beta^2}{1+\beta^2}
     \label{eq:Rbeta}
\end{equation}

Here, $m_p$, $\rho$, $U_{\rm gas}$ and $k_{\rm B}$ are the proton mass, GRMHD fluid density and internal energy, and finally, the Boltzmann constant. Thus, in this work, we use the black hole spin parameter $a_*$ and the temperature parameter $R_\mathrm{high}$ to characterize a library of raytraced images from \citep{Roelofs2021}. Understanding these images will give us valuable insights into the nature of space, time, and gravity, which are fundamental to our understanding of the universe. 
\\ \\
In recent years, deep-learning-based models have succeeded in various computer vision tasks. In particular, its applications in astrophysics have yielded impressive results. One  example is the use of computer vision techniques to study black holes, showcasing the versatility and effectiveness of this technology. \citep{deephorizon}, \citep{qiu} and \citep{tao} explored various machine learning methods to parameterize the M87* black hole. However, these frameworks rely heavily on having large training datasets and the high computational cost of GRMHD simulations leads to sparse training datasets. As a result, the development of accurate parameterization algorithms presents a significant challenge.
\\ \\
To overcome this problem, we propose a Conditional Progressive Generative Adversarial Network (CPGAN) to synthesize high-fidelity data that resemble the simulated images of GRMHD. Our architecture is capable of synthesizing images based on spin $a_*$ and $R_\mathrm{high}$. For a given $R_\mathrm{high}$, our model can generate black hole images with any spin in the continuous interval [-1, 1]. Considering the high cost associated with directly simulating finely-grained parameter values, for example, a spin of 0.64, the images generated by our network, if appropriately validated, prove to be valuable resources for more effectively sampling the parameter space in theoretical studies. Through our experiments, we demonstrate that our model can be used as a data augmentation tool for other deep learning algorithms, which can play a vital role in improving the parameterization of supermassive black holes, as well as facilitating a more comprehensive exploration of their characteristics and properties.
\\ \\
This paper is organized as follows. Section~\ref{sec:bg} presents the background information, including an overview of Generative Adversarial Networks (GANs), Wasserstein GANs, Progressive Growing GANs, and various evaluation metrics for GANs. Section~\ref{sec:data} focuses on the data used in our study, covering simulated data, dataset structure, and data augmentation techniques. In Section~\ref{sec:methods}, we describe our methodology, including the general approach, specific details of the GAN implementation, the parameterization network, and the model selection process. Section~\ref{sec:res} presents the results of our experiments, including an analysis of the generated samples and classification performance. Finally, in Section~\ref{sec:conc}, we summarize the conclusions drawn from our research and discuss future directions for further exploration in this field.

\section{Background} 
\label{sec:bg}
\subsection{Generative Adversarial Networks} 

Generative Adversarial Networks (GANs) \citep{gan} have shown remarkable results in image generation due to their ability to produce realistic and diverse images \citep{stylegan}, \citep{stylegan2}, \citep{stylegan3}, \citep{proGAN}, \citep{biggan}, \citep{cyclegan}.
\\ \\ 
GANs utilize game theory to approximate the probability distribution of a set of images by training two players, the generator and the discriminator, through a min-max game. The generator aims to generate realistic images to deceive the discriminator, while the discriminator strives to distinguish between real and generated images with high accuracy. Effectively, GANs learn a data distribution $P_{g}(x)$ that matches the real data distribution $P_{r}(x)$. Given real images $x_{r} \sim P_{r}$, GANs learn a generator network $G$ that transforms some input noise $z \sim P_{z}$ to a sample $x_{g} \sim P_{g}$. These generated samples are then passed on to a discriminator network $D$ whose role is to distinguish between samples from the real data distribution $P_{r}$ and the generated distribution $P_{g}$. This goal is achieved by optimizing the objective function:

\begin{equation}
    min_{G}max_{D}V(D,G) = E_{x\sim P_{r}}[logD(x)] + E_{z\sim noise}[log(1-D(G(z)))]
\end{equation}

\noindent
where $min_{G}max_{D}$ denote the minimization over the generator ($G$) and the maximization over the discriminator ($D$). The term $E_{x\sim P_{r}}[logD(x)]$ corresponds to the expected logarithm of the discriminator's output when evaluating real data samples $x$ drawn from $P_{r}$. Similarly, $E_{z\sim noise}[log(1-D(G(z)))]$ represents the expected logarithm of the discriminator's output when considering generated images $G(z)$ obtained by passing noise samples $z$ through the generator.
\\ \\
\noindent
Thus, one network, the generator, tries to create realistic data. The other network, the discriminator, tries to distinguish between the generated data and the real data. The generator keeps improving its output until it fools the discriminator into thinking that the generated data is real. The discriminator, on the other hand, keeps improving its ability to distinguish between the two types of data. This process continues until the generator can produce samples that are indistinguishable from the real data. 
\\ \\
In order to have control over the images generated by a GAN, the Auxiliary Conditional Generative Adversarial Network (ACGAN) was introduced \citep{acgan}. This framework enhances the GAN architecture by training the generator not only to deceive the discriminator but also to produce images with specific class labels. In ACGAN, the generator is conditioned on a class label vector ($y$), which is concatenated with the latent noise vector ($z$). By incorporating this class label information into the generator's input, the generator learns to generate images that correspond to a particular category or class.
\\ \\
Conversely, the discriminator network in ACGAN is modified to classify both real images and generated images into their respective classes. This modification enables the discriminator to not only differentiate between real and fake images but also discern the specific class to which an image belongs. By jointly training the generator and discriminator in this manner, ACGAN facilitates the generation of realistic images that possess the characteristics and attributes associated with a particular class.
\\\\
To further enhance ACGAN, we introduce a separate regressor in parallel with the discriminator, thereby separating the regression and discriminator tasks. This separation is necessary because the discriminator alone has limited flexibility in both, effectively discriminating between real and fake images, and capturing the spin of the black hole. The parallel operation allows the discriminator to focus on discrimination while the regressor specializes in the prediction of spin $a_*$. During training, the generator, discriminator, and classifier are optimized to minimize a composite loss function that includes the GAN loss as well as a regressor loss.

\subsection{Wasserstein GANs}
Wasserstein GANs (WGANs) \citep{wgan} are a variant of GANs specifically designed to improve the quality of generated samples. In WGANs, the Wasserstein distance, also known as the Earth Mover's distance, is used as the loss function for the GAN to train the generator and the discriminator, commonly called the critic. This distance measure provides a more informative and stable measure of the dissimilarity between the real data distribution ($P_r$) and the generated samples distribution ($P_g$).
\\ \\
To ensure the training process is stable, WGANs introduce a technique called weight clipping. Weight clipping involves constraining the weights of the critic network within a specific range. This constraint enforces that the critic function remains a 1-Lipschitz function, meaning its derivative is bounded by a constant value of 1. By enforcing the 1-Lipschitz condition through weight clipping, WGANs can alleviate issues such as vanishing or exploding gradients, thereby enabling the generator to learn the underlying data distribution more reliably, resulting in improved sample generation capabilities and reducing issues such as mode collapse.
\\ \\
WGAN-GP (WGAN with Gradient Penalty) \citep{wgan-gp} is an extension of WGAN that further stabilizes the training process. Instead of weight clipping, it adds a gradient penalty term to the loss function in order to enforce the Lipschitz constraint on the critic. The loss function for the critic in a WGAN-GP is given by:

\begin{equation}
    L_{G}= E_{x_{r}\sim P_{r}}[D(x_{r})] - E_{x_{g}\sim P_{g}}[D(x_{g})]
\end{equation}

\begin{equation}
    L_{D}= -L_{G} + \lambda E_{\hat{x}\sim{P_{\hat{x}}}}\left[(\|\nabla_{\hat{x}}D(\hat{x})\|_2 - 1)^2\right]
\end{equation}

\noindent
where $P_{\hat{x}}$ is the distribution implicitly defined by sampling uniformly along linear paths between points sampled from $P_{r}$ and $P_{g}$, and $\lambda$ is the penalty coefficient that controls the strength of the gradient regularization.

\subsection{Progressive Growing GANs}
Progressive Growing of GANs (PGGANs) \citep{proGAN} was designed to tackle challenges faced while generating high-resolution images. The main idea behind PGGANs is to start training GANs with low-resolution images and gradually increase the resolution of both the generator and discriminator over time.
\\ \\
PGGAN starts with a low-resolution version of the target image and gradually increases the resolution of the generated images as training progresses. This is achieved through progressive growth, where the generator and discriminator are trained on increasingly larger images at each stage. To facilitate this transition, a smooth alpha-fading technique is employed. This technique involves blending the parameters of the existing network with the newly introduced ones, ensuring a seamless transition in the network's architecture. The weights from the previous stage are used as initialization for the next stage, allowing the network to build on its knowledge at each stage and helping prevent the generator from collapsing into a single mode.
\\ \\
Another key aspect of PGGAN is the use of a multi-scale discriminator, which is capable of processing images at multiple resolutions and provides more detailed feedback to the generator about the quality of the generated samples. This can help the generator produce more realistic images and improve the overall performance of the GAN.

\subsection{Evaluation of GANs}
The loss functions used in GANs effectively address the adversarial problem by optimizing the generator-discriminator dynamics. However, solely using the loss function as an evaluation metric is inadequate for assessing the quality of the generated samples. This is because the loss function is designed specifically for the generator trained within the GAN framework and may not generalize well to other datasets or downstream tasks. In other words, the evaluation based solely on the loss function does not capture the full range of performance characteristics. Therefore, it is crucial to incorporate additional evaluation metrics that provide a more comprehensive assessment of the generated samples.
\\ \\
To assess the performance of GANs, it is necessary to measure the distance between the probability distributions of real ($P_{r}$) and generated ($P_{g}$) data. However, when dealing with high-dimensional distributions, devising a suitable metric is challenging and remains an open problem. Furthermore, the evaluation metrics for GANs must effectively measure both fidelity and diversity. Fidelity is a measure of similarity between the characteristics of the real and generated samples, while diversity measures the variance of the generated samples. In other words, fidelity is an assessment of realism in the generated samples, while diversity assesses how much of the real data distribution is captured by the generated samples.

\subsubsection{Frechét Distance (FID)}
Frechet Distance introduced by \citep{fid} is a measure of the distance between two probability distributions. To compute the frechet distance, we first need to extract feature vectors by mapping the samples $x_{r}$ or $x_{g}$  to an intermediate feature space. This mapping is generally performed using a convolutional neural network (usually InceptionV3 \citep{inceptionv3} ). Once we have the feature vectors, we can compute the FID as follows:

\begin{equation}
    FID = \lVert\mu_{g} - \mu_{r}\rVert^2 + \operatorname{Tr}(C_r + C_g - 2\sqrt{C_g C_r})
\end{equation}
\noindent
where $\mu_g$ and $\mu_r$ are the means of the feature vectors for the generated and reference samples, respectively, $C_g$ and $C_r$ are the covariance matrices of the feature vectors for the generated and reference samples, respectively, and $Tr(A)$ is the trace of matrix $A$.

\subsubsection{Precision and Recall}
The use of precision and recall as an evaluation metric, as suggested by \citep{precisionrecall}, provides us with separate measurements for fidelity and diversity, allowing for a more comprehensive evaluation of the generated samples.
\\ \\
Precision assesses the fidelity of the image by measuring the fraction of generated samples that belong to the real distribution. On the other hand, recall assesses the diversity of the image by measuring the fraction of real samples that are correctly captured by the generated distribution. 

\subsubsection{TSTR and TRTR}
\label{sec:tstr_trtr}
An indirect approach to measuring the similarity between the real ($P_{r}$) and fake distributions ($P_{g}$) is to compare the performance of the generated samples on downstream tasks. If a generative model can successfully learn $P_{r}$, then the data drawn from either the real or generated distributions will yield similar results on the appropriately selected downstream task.
\\ \\
In our work, we define the downstream objective as the parametrization of the target image using a regressor. The $R^2$ score, also known as the coefficient of determination, measures the proportion of the variance in the target variable, in our case the spin $a_{*}$, that can be explained by the regressor. It indicates how well the regression model fits the observed data, with values closer to 1 indicating a better fit. By comparing the $R2$ scores obtained in different scenarios, such as training on synthetic or GRMHD data and testing on GRMHD data, we can assess the similarity between the performance of the generative model and real data on the downstream task of image parametrization.. Performance in these two scenarios, train on synthetic test on real (TSTR), and train on real test on real (TRTR) \citep{yang} are expected to be similar.  Additionally, we measure performance when the parameterization network is trained on an augmented dataset, consisting of a mix of real and generated samples, and tested on the real dataset (T(S+R)TR).

\section{Data}
\label{sec:data}
In addition to providing the event-horizon scale images of M87*, Event Horizon Telescope Collaboration \citep[EHTC][]{eht_m87_p1,eht_m87_p2,eht_m87_p3,eht_m87_p4,eht_m87_p5,eht_m87_p6} successfully modeled the appearance of M87* using GRMHD simulations. Each simulation was used to describe several different physical scenarios, and each scenario was then used to generate several simulated images.
\\ \\
Since the GRMHD-simulated data was generated based on a range of parameters and labeled accordingly, it was a sensible choice for use as the training data. The data used in this project are labeled based on four key parameters: $a_*$ (spin), $R_\mathrm{high}$, frequency and frame.
\\ \\
Using these labeled data, a generative model can learn to identify features that characterize a given label and generate new data points that are virtually indistinguishable to simulation experts from the GRMHD simulated data, allowing for the creation of a larger and more diverse dataset for parameterization algorithms and detailed study.
\subsection{Simulated Data}
\begin{figure}
    \includegraphics[width=\columnwidth]{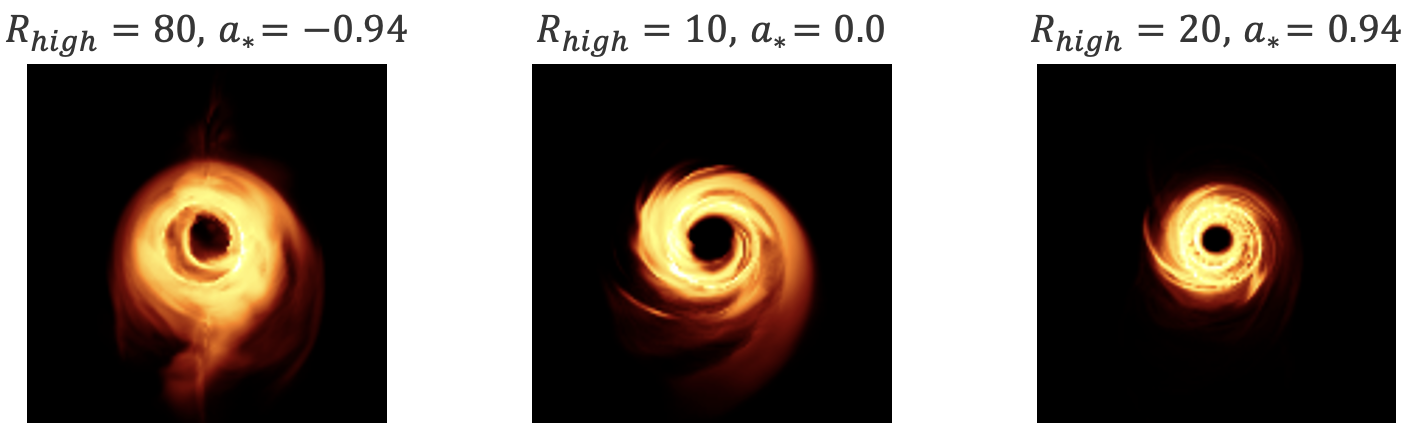}
    \caption{Three example snapshots from our image library corresponding to different spin parameters and accretion flows.}
    \label{fig:GRMHD}
\end{figure}

In the EHT collaboration, the accretion flow of the M87* was modeled using GRMHD simulations, which was found to successfully describe a turbulent, hot, magnetized disk orbiting a Kerr black hole \citep[EHTC][]{eht_m87_p1,eht_m87_p2,eht_m87_p3,eht_m87_p4,eht_m87_p5,eht_m87_p6}. To further explore the imaging results, the researchers created a Simulation Library \citep[EHTC][]{eht_m87_p1,eht_m87_p2,eht_m87_p3,eht_m87_p4,eht_m87_p5,eht_m87_p6} by fitting GRMHD models to the image of M87*. They then used the elements of this Simulation Library to create an Image Library \citep[EHTC][]{eht_m87_p1,eht_m87_p2,eht_m87_p3,eht_m87_p4,eht_m87_p5,eht_m87_p6} through GRRT simulations.
\\ \\
A typical GRMHD simulation in the library is characterized by two parameters: (1) the dimensionless spin $a_* \equiv Jc/GM^2$, where $J$ and $M$ are the spin angular momentum and mass of the black hole, respectively, and (2) the net dimensionless magnetic flux over the event horizon $\phi = \Phi/(\dot{M}R_{g}^{2})^{1/2}$ , where $\Phi$ and $\dot{M}$ are the magnetic flux and mass flux (or accretion rate) across the horizon, respectively \citep[EHTC][]{eht_m87_p1,eht_m87_p2,eht_m87_p3,eht_m87_p4,eht_m87_p5,eht_m87_p6}. Based on these parameters, the models are either “Standard and Normal Evolution” (SANE) if $\phi \sim 1$ or “Magnetically Arrested Disk” (MAD) if $\phi \sim 15$. The models also include five $a_*$ values that make the accretion disk either “prograde” if the motion of the accretion disk with respect to its spin axis is in the same direction of black hole spin ( $a_{*} \geq 0$) or “retrograde” if the motion is in the opposite direction ($a_{*} < 0$). By varying $a_{*}$ and $\phi$, they have thoroughly explored the physical properties of magnetized accretion flows onto Kerr black holes.
\\ \\
Once a simulation is completed, the image library is then constructed on the basis of these GRMHD models by performing general relativistic ray-tracing (GRRT) simulations. A GRRT simulation is characterized by the properties of the fluid, the emission and absorption coefficients, the inclination angle between the angular-momentum vector of the accretion-flow and the line of sight, the position angle, the mass of the black hole, and the distance from the observer \citep[EHTC][]{eht_m87_p1,eht_m87_p2,eht_m87_p3,eht_m87_p4,eht_m87_p5,eht_m87_p6}. Here, we created a new image library consisting of SANE GRMHD models generated with the GRMHD code \texttt{BHAC} \citep{Porth2017} and the GRRT images performed by the radiative transfer code \texttt{BHOSS} code \citep{Younsi2012,Younsi2020}. During the GRRT we assumed a thermal electron distribution and seven values of the temperature ratio of electrons to protons, parameterized by $R_\mathrm{high}$, one mass, and two inclinations \citep[for more details on the GRMHD and GRRT simulations see][]{Roelofs2021,Fromm2022}.\\
\\ \\
Depending on the frequency mode determined at the time of GRRT imaging, the features exhibited by a black hole can vary, as the change in energy distribution can affect the radiation spectra. A higher frequency corresponds to higher resolution and a deeper view of the accretion flow, since the angular resolution is proportional to the observing wavelength ($\lambda$) and the maximum projected baseline length between the telescopes in the array $L$ \citep{thompson}. In this project, we utilized SANE images operating at a single frequency mode of 230 GHz, which corresponds to the current operating frequency of the EHT. 
\\ \\
Therefore, the simulated images from the Image Library, which fitted closure phases and amplitudes of the April 11 \citep[EHTC][]{eht_m87_p1,eht_m87_p2,eht_m87_p3,eht_m87_p4,eht_m87_p5,eht_m87_p6} data best as seen in Fig.~\ref{fig:GRMHD} can correspond to different spins and different flow types. 

\subsection{Dataset Structure}
The GRMHD simulated images are characterized by 2 parameters: (1) $a_{*}$, consisting of five different values, and (2) $R_\mathrm{high}$, consisting of seven different values, which correspond to a total of 35 unique parameter combinations. Furthermore, each simulation generates a series of frames that represent the rotating motion of the black hole.
\\ \\
In our study, we have created a dataset (Table~\ref{table:dataset}) by treating all 101 frames of each simulation as individual static images, allowing for a more detailed analysis of the image properties and facilitating machine learning-based parameter estimation algorithms. It is also worth noting that the time spacing between the frames corresponds to 3.5 days. While our study focuses on image generation, generating videos is an area for further investigation.

\begin{table}
    \centering
    \caption{Dataset structure.}
    \begin{tabular}{cc}
    \hline
     Parameter & Values \\
    \hline
     Accretion & SANE\\

     Frequency & 230 GHz, 345 GHz \\

     $a_*$ & $-0.94$, $-0.50$, $0.00$, $0.50$, $0.94$ \\

     $R_\mathrm{high}$ & 1, 5, 10, 20, 40, 80, 160 \\

     Frames & 1, 2, 3, 4, ..., 100, 101 \\

     \textbf{Inclination Angle} & \textbf{163 deg}\\

    \hline
    \end{tabular}
    
    \label{table:dataset}
\end{table}

\subsection{Data Augmentation}
Data augmentation is a process of creating additional training data by applying various transformations to existing data. In the case of image data, these transformations can include flipping, rotating, zooming, and cropping, among others. While data augmentation is commonly used to enforce symmetries in the learned function of a neural network, it is also widely used to increase the fidelity of the function by enriching the dataset. Increasing the size and diversity of the training dataset through data augmentation can help the machine learning model learn more robust features and reduce overfitting by generalizing better to unseen data \citep[][]{shortkhosh,perez}.
\\ \\
In this paper, we performed these data augmentation transformations on the existing primary data source from GRMHD simulations. Specifically, the primary dataset was rotated in 90-degree steps, quadrupling the training data. This technique allowed the GAN to be trained on a larger dataset, which ultimately led to better model performance and the generation of more realistic simulated images.
\\ \\
It is important to note that this transformation did not alter the underlying physics in any way, as the position angle is arbitrary.  In particular, the direction of the brightness asymmetry of the image depends on the projection of the black hole's spin axis onto the sky plane, which is not of physical significance.  On the contrary, this augmentation made the model robust against different rotations of the images as a result of the observer's line of sight.  It is important for the model not to use the direction of the brightness asymmetry to distinguish retrograde from prograde models, for example.

\section{Methodology}
\label{sec:methods}
\subsection{General Description}
\begin{figure}
    \includegraphics[width=\columnwidth]{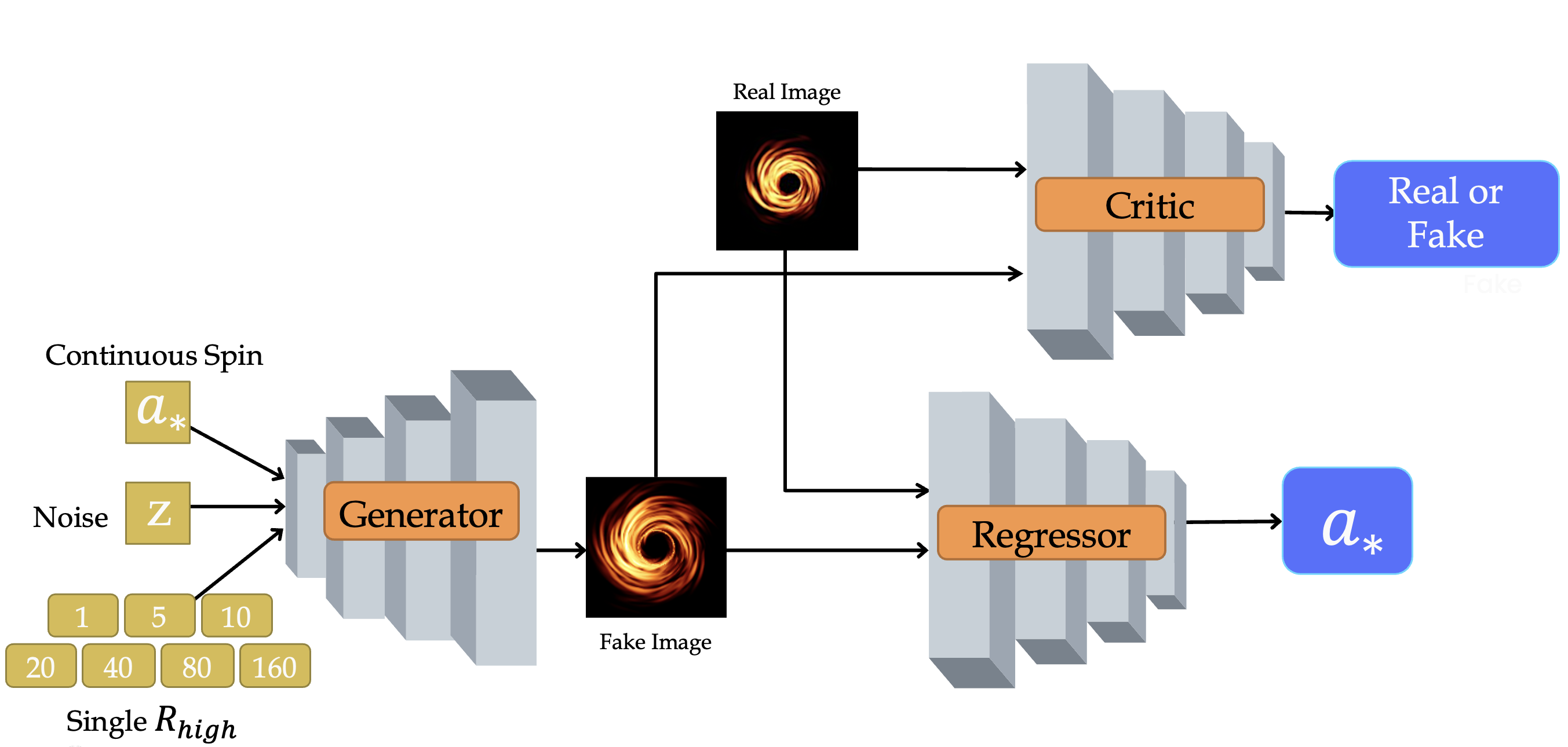}
    \caption{Architecture of our CPGAN. It takes spin $a_{*}$ and $R_\mathrm{high}$ as inputs and generates 128x128 pixel images.}
    \label{fig:GAN_Details}
\end{figure}

We propose a data augmentation methodology that involves a Conditional Progressive Generative Adversarial Network (CPGAN) model designed to generate new images of black holes based on specific spin and electron distribution parameters. In this framework, the CPGAN consists of three convolutional neural networks: a generator that produces the synthetic images based on $a_*$ and $R_\mathrm{high}$, a critic that differentiates between real and synthetic images, and an auxiliary regressor that predicts the spin of a given image as shown in Fig.~\ref{fig:GAN_Details}. Our proposed CPGAN model can generate new images for any $a_{*}$ in the $[-1, 1]$ range. Additionally, by training one CPGAN for every $R_\mathrm{high}$, we can ensure that our model performs optimally for the entire range of $R_\mathrm{high}$ and $a_{*}$ values, resulting in a more robust and accurate approach. This setup helps us increase the size of the training dataset, allowing for more accurate parameterization of a black hole image. The details of how the conditional progressive GAN is built, including the architecture and loss function, will be explained in Section~\ref{sec:gandetails}.
\\ \\
To evaluate the effectiveness of our proposed methodology, we utilize two evaluation metrics: TSTR and TRTR, which were previously discussed in Section~\ref{sec:tstr_trtr}. To compute these metrics, we employ a convolutional neural network (CNN), as proposed in \citep{tao}. The CNN takes as input either the GRMHD-simulated images (for TRTR) or the generated images (for TSTR) and predicts the corresponding values of $a_{*}$. Further details of the CNN architecture and training process will be explained in Section~\ref{sec:param_nw}. 

\begin{figure}
    \includegraphics[width=\columnwidth]{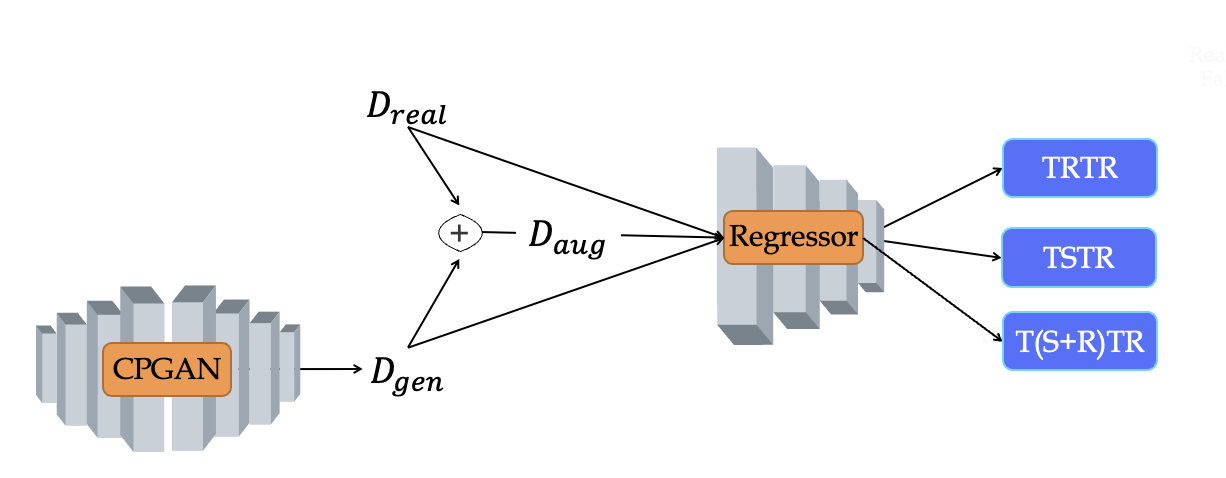}
    \caption{Diagram of the methodology.}
    \label{fig:General}
\end{figure}
\noindent
\\ \\
Fig.~\ref{fig:General} provides a summary of the proposed methodology, illustrating the flow of data and models involved in the process.

\subsection{GAN details}
\label{sec:gandetails}
In this section, we provide a detailed explanation of the GAN architecture used in our proposed methodology. Our GAN framework consists of three distinct networks: generator, critic, and regressor. 
\\ \\
Following the conditional GAN \citep{cgan} architecture, the generator takes as input a conditioning variable $a_{*}$ and the noise vector. The input is a concatenated vector $\bar{z} = [z, a_{*}]$, where  $a_{*}$ is sampled from five discrete spins in our dataset and $z$ is a noise vector such that $z \in \mathbb{R}^l \sim \mathcal{N}(0, I)$. Here, we choose $l$ to be a 512-dimensional noise vector. Table~\ref{table:gentable} shows the detailed generator architecture used to generate a 128 x 128 pixel image of a black hole. 
\\ \\
The critic network takes an image as its input, either belonging to the real dataset or an image generated by the generator, and is trained to distinguish between real and generated images. Here, the output of the critic is a scalar value that represents the realness of the image. The critic is trained to output high values for real images and low values for generated images. In addition, our architecture incorporates a minibatch standard deviation layer, as described in Table~\ref{table:disctable}, which calculates the standard deviation of features within a mini-batch of data. This layer introduces diversity into the generated samples by considering the measure of diversity, allowing the generator to produce more varied outputs.
\\ \\
The regressor network takes the generated image as input and outputs its prediction of spin ($a_{*}$) as seen in Fig.~\ref{fig:GAN_Details}. Our goal with the regressor network is to accurately predict the value of spin for each generated image. The gradients from the regressor help in reinforcing the connection between the generated image and its associated spin. All three of our networks using the Leakly ReLU (LReLU) activation function. It is an extension of the traditional ReLU (Rectified Linear Unit) activation function. In a ReLU, when the input is greater than zero, it returns the input value, and when it's less than zero, it returns zero. In contrast, LReLU allows a small gradient for negative values, typically by multiplying the input by a small positive constant. This slight "leak" for negative inputs helps prevent the vanishing gradient problem and enables the network to learn more effectively.
\\ \\
The loss function used to train this GAN is as follows:
\begin{equation}
    L_{G}= E_{x_{r}\sim P_{r}}[D(x_{r})] - E_{x_{g}\sim P_{g}}[D(x_{g})] + \sum_{i=1}^{N}(a_{*}-R(x_{g}))^2
\end{equation}

\begin{equation}
    L_{D}= -L_{G} + \lambda E_{\hat{x}\sim{P_{\hat{x}}}}\left[(\|\nabla_{\hat{x}}D(\hat{x})\|_2 - 1)^2\right]
\end{equation}

\begin{equation}
    L_{R}= \sum_{i=1}^{N}(a_{*}-R(x_{g}))^2 + \sum_{i=1}^{N}(a_{*}-R(x_{r}))^2
\end{equation}
where $P_{\hat{x}}$ is the distribution implicitly defined by sampling uniformly along linear paths between points sampled from $P_{r}$ and $P_{g}$, $x_{r}$ is a data point sampled from $P_{r}$, $x_{g}$ is a data point sampled from $P_{g}$, $\lambda$ is the penalty coefficient that controls the strength of the gradient regularization and $a_{*}$ is all of the spins from the real dataset.
\\ \\
To enhance the quality of the generated images, we employed a progressive training strategy, training all three networks while gradually increasing the image resolution from 4 x 4 pixels to 128 x 128 pixels. This approach allows the generator to learn the fundamental structure of the images at lower resolutions before capturing finer details. It is worth noting that the GRMHD simulation images themselves were not generated at different resolutions; however, to ensure a fair comparison between the synthetic and actual data, we applied a downsampling process to the real images. During progressive training, we also employed a smooth alpha-fading technique to blend the images generated by different network configurations. This technique mitigates artifacts that can arise when the generator abruptly transitions to a new configuration. By smoothly blending the outputs, the network produces visually coherent images at each training step, allowing us to evaluate the fidelity and realism of the generated images throughout the training process.
\\ \\
\begin{table}
    \centering
    \caption{Generator architecture.}
    \begin{tabular}{c c c}
      \hline
      \textbf{Generator} & Activation & Output Shape \\
      \hline
      Latent Vector & - & 1 x 1 x 512\\
      $a_*$ & - & 1 x 1 x 1\\
      Concatenate & - & 1 x 1 x 513\\
      Dense & LReLU & 1 x 1 x 8192\\
      Reshape & - & 4 x 4 x 512\\
      Conv 4 x 4 & LReLU & 4 x 4 x 512\\
      Conv 3 x 3 & LReLU & 4 x 4 x 512\\
    
      Upsample & - & 8 x 8 x 512\\
      Conv 3 x 3 & LReLU & 8 x 8 x 512\\
      Conv 3 x 3 & LReLU & 8 x 8 x 256\\
  
      Upsample & - & 16 x 16 x 256\\
      Conv 3 x 3 & LReLU & 16 x 16 x 256\\
      Conv 3 x 3 & LReLU & 16 x 16 x 128\\
     
      Upsample & - & 32 x 32 x 128\\
      Conv 3 x 3 & LReLU & 32 x 32 x 128\\
      Conv 3 x 3 & LReLU & 32 x 32 x 64\\
  
      Upsample & - & 64 x 64 x 64\\
      Conv 3 x 3 & LReLU & 64 x 64 x 64\\
      Conv 3 x 3 & LReLU & 64 x 64 x 32\\
   
      Upsample & - & 128 x 128 x 32\\
      Conv 3 x 3 & LReLU & 128 x 128 x 32\\
      Conv 3 x 3 & LReLU & 128 x 128 x 16\\
      Conv 1 x 1 & tanh & 128 x 128 x 2\\
      \hline
    \end{tabular}
    
    \label{table:gentable}
\end{table}

\begin{table}
    \centering
    \caption{Critic architecture.}
    \begin{tabular}{c c c}
      \hline
      \textbf{Critic} & Activation & Output Shape \\
      \hline
      Input Image & - & 128 x 128 x 2\\
      Conv 1 x 1 & LReLU & 128 x 128 x 16\\
      Conv 3 x 3 & LReLU & 128 x 128 x 16\\
      Conv 3 x 3 & LReLU & 128 x 128 x 32\\
      Downsample & - & 64 x 64 x 32\\
      
      Conv 3 x 3 & LReLU & 64 x 64 x 32\\
      Conv 3 x 3 & LReLU & 128 x 128 x 64\\
      Downsample & - & 32 x 32 x 64\\
   
      Conv 3 x 3 & LReLU & 32 x 32 x 64\\
      Conv 3 x 3 & LReLU & 32 x 32 x 128\\
      Downsample & - & 16 x 16 x 128\\

      Conv 3 x 3 & LReLU & 16 x 16 x 128\\
      Conv 3 x 3 & LReLU & 16 x 16 x 256\\
      Downsample & - & 8 x 8 x 256\\

      Conv 3 x 3 & LReLU & 8 x 8 x 256\\
      Conv 3 x 3 & LReLU & 8 x 8 x 512\\
      Downsample & - & 4 x 4 x 512\\
  
      Minibatch stddev & - & 4 x 4 x 513\\
      Conv 3 x 3 & LReLU & 4 x 4 x 512\\
      Conv 4 x 4 & LReLU & 1 x 1 x 512\\
      Dense & Linear & 1 x 1 x 1\\
      \hline

    \end{tabular}
    
    \label{table:disctable}
\end{table}

\begin{table}
    \centering
    \caption{Regressor architecture.}
    \begin{tabular}{c c c}
      \hline
      \textbf{Regressor} & Activation & Output Shape \\
      \hline
      Input Image & - & 128 x 128 x 2\\
      Conv 3 x 3 & LReLU & 128 x 128 x 10\\
      Downsample & - & 64 x 64 x 10\\

      Conv 3 x 3 & LReLU & 64 x 64 x 20\\
      Downsample & - & 32 x 32 x 20\\

      Conv 3 x 3 & LReLU & 32 x 32 x 50\\
      Downsample & - & 16 x 16 x 50\\

      Conv 3 x 3 & LReLU & 16 x 16 x 50\\
      Downsample & - & 8 x 8 x 50\\

      Conv 3 x 3 & LReLU & 8 x 8 x 50\\
      Downsample & - & 4 x 4 x 50\\

      Conv 3 x 3 & LReLU & 4 x 4 x 50\\
      Downsample & - & 1 x 1 x 50\\
      Dense & LReLU & 1 x 1 x 16\\
      Dense & Linear & 1 x 1 x 1\\
      \hline
    \end{tabular}
    
    \label{table:regtable}
\end{table}

\noindent 
We conducted various experiments to achieve optimal results, exploring different configurations. Through these experiments, we discovered that a training approach involving alternating between the generator and critic networks yielded favorable outcomes. Specifically, we trained the critic network five times for every training iteration of the generator. Additionally, we utilized the WGAN-GP loss \citep{wgan}, which further contributed to the superior performance of our Generative Adversarial Network (GAN) in generating images of black holes. We added a mean squared error regularization term to train the regressor network, which measures the average squared difference between the predicted spin values and the true spin values.
\\ \\
It is worth noting that we are training the GAN on discrete labels, i.e., five discrete values of $a_{*}$ and seven discrete values of $R_\mathrm{high}$. However, our GAN can generate images for any spin value between -1 and 1, given a discrete $R_\mathrm{high}$ value. This allows us to generate a more diverse set of black hole images.
\\ \\
Overall, our proposed GAN framework as seen in Fig.~\ref{fig:GAN_Details} provides a powerful tool for generating new images of a black hole based on specific spin and electron distribution parameters. In Section~\ref{sec:res}, we will explain in detail how we evaluate the performance of our proposed methodology.

\subsection{Parameterization Network}
\label{sec:param_nw}
\begin{figure}
    \includegraphics[width=\columnwidth]{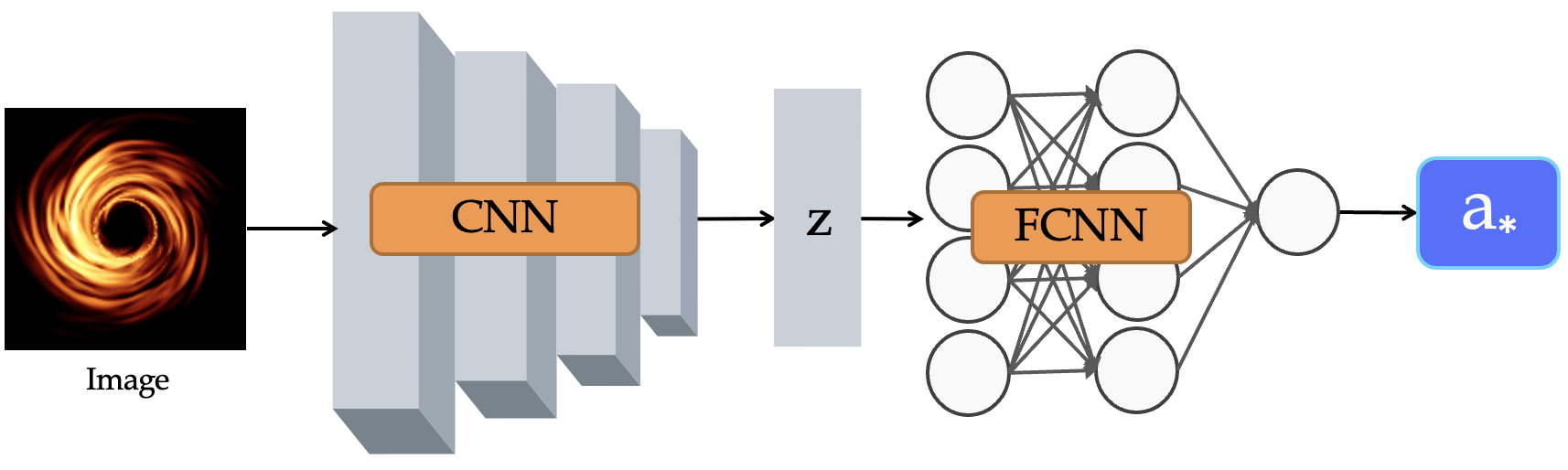}
    \caption{\citet{tao}'s architecture of the CNN regressor, which takes in images with format 128 x 128 x 2. The convolutional layers transform the original image into a vector $z$, where $z \in  \mathbb{R}^50$. Eventually, the fully-connected neural network (FCNN) projects $z$ into a single variable output $a_{*}$.}
    \label{fig:Tao_Classifier}
\end{figure}

To evaluate the performance of our generated model, we use a downstream parameterization task. We use a convolutional neural network (CNN)  since they are translationally invariant, and the neurons in each layer are locally connected. This guarantees that the position of black holes in the input images is not restricted to the center and that the spatial characteristics of the black hole image, such as asymmetry in the emission ring can be effectively captured. For our project, we use the CNN regressor architecture proposed by \citep{tao}, since it has been shown to be effective in relating spatial features in black hole images to parameters such as $a_{\star}$ and $R_\mathrm{high}$. The regressor consists of CNN layers that transform the input image to a vector (z) and a fully connected layer that maps this vector to a single neuron that estimates $a_{*}$. Fig.~\ref{fig:Tao_Classifier} provides a visual representation of this process. The architecture used can be seen in the Appendix.
\\ \\ 
We train this network on both synthetic images (TSTR), real images (TRTR), and an augmented dataset consisting of real and fake images (T(S+R)TR). We then compute the $R^2$ score in each of the three scenarios. A detailed overview of the results is presented in Section~\ref{sec:res}.

\subsection{Model Selection}
To select the best generator model for our proposed methodology, we utilized the Fréchet Inception Distance (FID) metric, which is commonly used to evaluate the similarity between the distribution of real images and generated images. The lower the FID score, the closer the generated images are to the real images.
\\ \\
We plotted the FID score against the number of training epochs to track the performance of the generator during training. We identified the epoch at which the FID score had reached its minimum value and selected that as the stopping point for training our generator model. The results of computing the FID for the GANs trained with different $R_\mathrm{high}$ values.
\\ \\
After selecting the best generator model, we used it in our parameterization network to estimate $a_{*}$ and $R_\mathrm{high}$ of the black hole image. We evaluated the accuracy of our parameterization network using three different evaluation methods: Train Real Test Real (TRTR), Train Synthetic Test Real  (TSTR), and Train Real + Synthetic Test Real (T(S+R)TR).

\section{Results}
\label{sec:res}
\subsection{Generated Samples}
\begin{figure}
    \centering
    \includegraphics[scale=0.5]{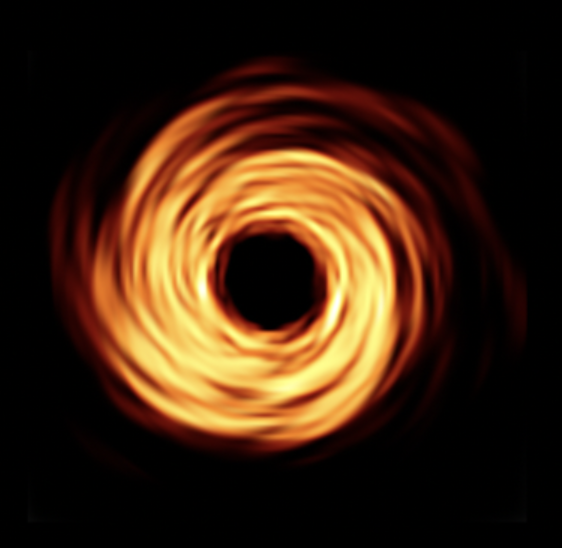}
    \caption{Generated sample for $a_{*}$ = 0.0 and $R_\mathrm{high}$ = 1}
    \label{fig:Generated_Images_one}
\end{figure}

\begin{figure}
    \includegraphics[width=\columnwidth]{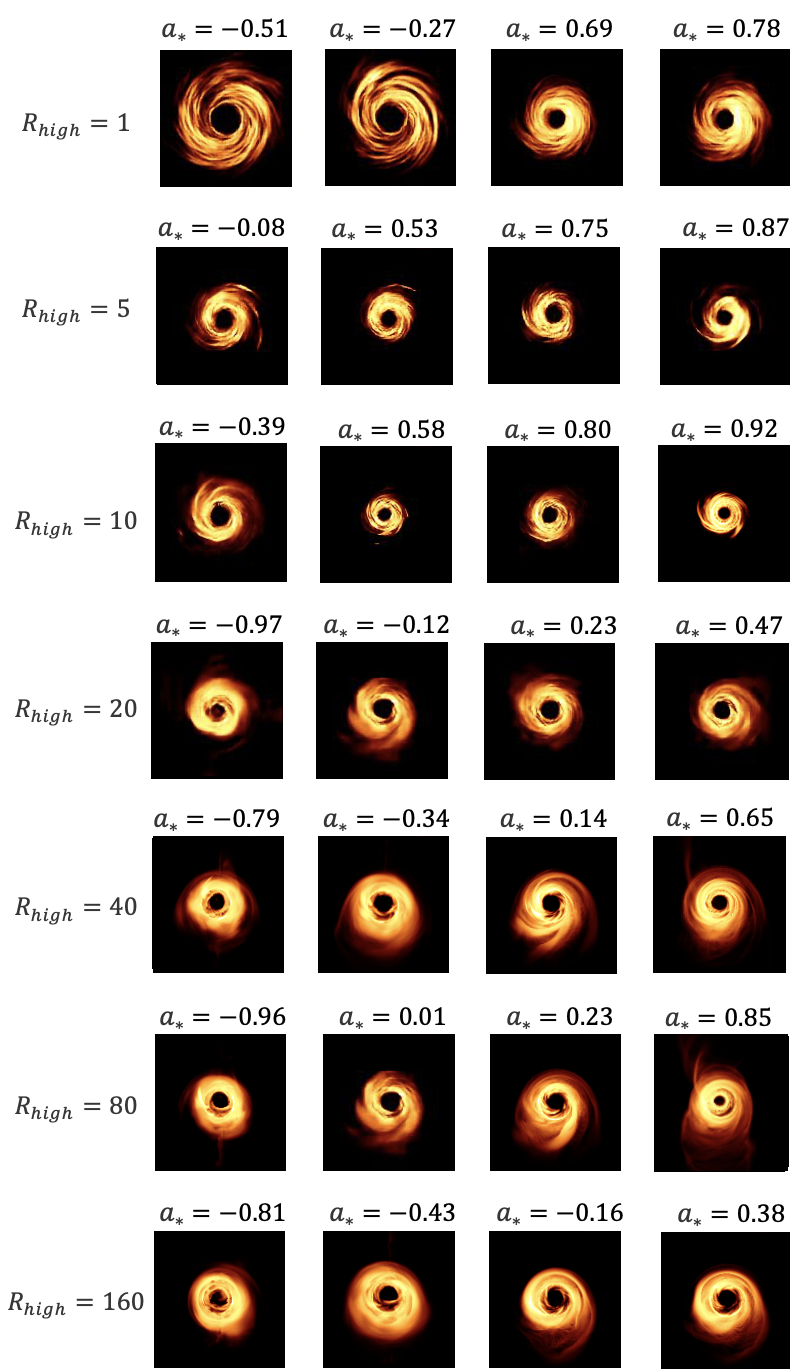}
    \caption{Generated samples for a given $a_{*}$ and $R_\mathrm{high}$.}
    \label{fig:Generated_Images}
\end{figure}

\noindent
Our proposed CPGAN  model can successfully generate black hole images for any spin between -1 and 1 and a given value of the seven $R_\mathrm{high}$. It is clear from the images that our model is able to generate high-quality samples that capture the characteristic features of the real images. These images are presented in Fig.~\ref{fig:Generated_Images}.
\\ \\
The generator successfully captures the photon ring and black hole shadow across different spin and $R_\mathrm{high}$ values. We observed that increasing the value of $R_\mathrm{high}$ leads to a reduction in the central brightness depression. This effect can be attributed to changes in the electron temperature distribution within the disk and jet. Specifically, increasing $R_\mathrm{high}$ decreases the temperature in the disk regions ($\beta>0$) while in the low $\beta$ regions $(\beta<10^{-2})$ the temperature is not changed. Consequently, more emission is generated in the jet, which is positioned in front of the shadow. The emitted radiation, primarily from the forward jet, blends over the flux depression, resulting in a reduction in the brightness depression.
\\ \\
Fig.~\ref{fig:Generated_Images} illustrates that higher $R_\mathrm{high}$ values introduce additional structure within the inner ring of the generated images. Conversely, lower $R_\mathrm{high}$ values do not exhibit significant inner ring structure, although negative spins display prominent spirals attributed to plasma accretion. These observations align with the findings observed in the GRMHD data, ensuring consistent and reliable results.
\\ \\
Overall, the generated samples indicate that our proposed methodology is able to generate high-quality images of a black hole for a wide range of spin and $R_\mathrm{high}$ values, which will enable more accurate parameterization of other black hole images.

\subsection{Classification}
\noindent
We then evaluated the performance of our proposed methodology using the TSTR, TRTR, and augmented T(S+R)TR $R^2$ scores. The R-squared ($R^2$) score measures the proportion of the variance in the predicted parameter values that can be explained by the actual parameter values. To compute the TRTR score, we trained and tested the regressor model using real data from the GRMHD simulations. Similarly, for the TSTR score, we generated synthetic samples using our proposed methodology and trained the regressor model using this training set. We then tested the regression model on the real testing set. Finally, to compute the T(S+R)TR, we combined the real and synthetic datasets and trained our parameterization network on this augmented dataset. We then tested the regression model on the real testing set.
\\ \\
Table~\ref{table:r2scores} displays the $R^2$ scores for the TSTR, TRTR, and augmented T(S+R)TR experiments, where values closer to one indicate higher accuracy. We observe that the $R^2$ score on the TSTR dataset is higher compared to the TRTR dataset. More importantly, however, we found that the $R^2$ score on the TS+RTR dataset was significantly higher compared to both TSTR and TRTR datasets. Here, it is worth noting that the $R^2$ score can differ when calculated on different subsets of the test data. In our study, we consistently found that the $R^2$ scores for TSTR and T(S+R)TR were higher for every test dataset compared to TRTR. The results presented in this section represent the highest $R^2$ scores obtained across all the test sets. We attribute this improvement in performance to the data augmentation provided by the GAN, which increases the size of the training dataset and allows for more accurate parameterization of black hole images. This demonstrates the effectiveness of our proposed data augmentation methodology using a conditional progressive GAN and its ability to generalize well to the real data. As shown in Table~\ref{table:r2scores}, the R2 scores performed well, indicating that our generation network identified patterns in the spin and the images.
\\ \\
Furthermore, we trained our generator using only four $a_{*}$ (-0.94, -0.50, 0.50, 0.94) and seven $R_\mathrm{high}$ values (1, 5, 10, 20, 40, 80, 160), leaving out one value ($a_{*}=0.00$) to understand if our model could generalize well to continuous spin values. In this case, we achieved a T(S+R')TR of 0.984, where R' indicates the dataset without one spin value. 
\\ \\
\begin{table}
    \centering
    \caption{Classification $R^{2}$ scores on different train and test datasets, where values closer to one correspond to a better match.}
    \begin{tabular}{ c  c c  c }
    \hline
     $R_\mathrm{high}$ & TRTR & TSTR & T(S+R)TR    \\
     \hline
     1 &  0.975 &  0.978 &  0.985 \\

     5 &  0.926 &  0.909 &  0.931\\

     10 &  0.917 &  0.935 &  0.943\\

     20 &  0.988 &  0.879 &  0.915\\

     40 &  0.976 &  0.903 &  0.981\\

     80 &  0.905 &  0.971 &  0.963\\

     160 &  0.938 &  0.910 &  0.950\\
    \hline
    \end{tabular}
    
    \label{table:r2scores}
\end{table}

\noindent 
In summary, the results of our evaluation indicate that our proposed methodology can help in the accurate parameterization of a black hole image for a wide range of spin and $R_\mathrm{high}$ values, and is able to generalize well to real data.

\section{Conclusions \& Future Work} 
\label{sec:conc}
In this study, we present a data augmentation methodology based on Conditional Progressive Generative Adversarial Networks (CPGAN) to generate diverse black hole (BH) images. The methodology takes into account variations in spin ($a_*$) and $R_\mathrm{high}$ prescriptions, resulting in a valuable resource for training deep learning algorithms to accurately estimate black hole parameters from observational data. 
\\ \\
Our proposed model is capable of generating BH images for any spin value within the range of [-1, 1], given a single $R_\mathrm{high}$. To assess the effectiveness of our approach, we employed a convolutional neural network (CNN) to predict the BH spin using both the General Relativistic Magnetohydrodynamics (GRMHD) images and the images generated by our CPGAN model. The results demonstrate a significant improvement in performance when training is performed using the augmented dataset. This improvement is indicated by a high $R^2$ score, highlighting the enhanced accuracy in BH spin prediction. 
\\ \\
Consequently, we propose that leveraging GANs for black hole image generation can be a cost-effective method to reliably augment training datasets for other parametrization algorithms. This paper contributes to the field of image-generation techniques for black holes and highlights the potential of deep learning algorithms in estimating important physical parameters from observational data. By expanding the training dataset with diverse and realistic BH images, our methodology enables more accurate parameterization and estimation and can lead to improved insights into the nature of black holes.
\\ \\
In our future work, we aim to expand the scope of our research beyond M87* and explore the application of our model to other black hole images, including the recently released images of Sagittarius A*. Additionally, while our primary focus was on SANE accretion models, we also intend to include MAD models in our investigations.
\\ \\
Furthermore, we plan to explore the impact of incorporating additional physical parameters, such as the inclination angle, as input to our model, to enhance the realism and diversity of the resulting images.
\\ \\
In order to improve the quality and stability of image generation, we will consider alternative architectures, specifically diffusion models, which have demonstrated the ability to produce high-quality images without suffering from training instability and mode collapse \citep{rombach}. By integrating these advanced networks into our dataset, we anticipate developing more sophisticated generative models tailored for the task of astronomical data generation \citep{mudur, doorenbos}. 

\section*{Acknowledgements}
CMF is supported by the DFG research grant “Jet physics on horizon scales and beyond” (Grant No. FR 4069/2-1). YM is supported by the National Natural Science Foundation of China (Grant No. 12273022) and Shanghai target program of basic research for international scientists (Grant No. 22JC1410600). The GRMHD simulations and GRRT calculations were performed on LOEWE at the CSC-Frankfurt, Iboga at ITP Frankfurt and Pi2.0 and Siyuan Mark-I at Shanghai Jiao Tong University and on MISTRAL at the University of W\"urzburg. Razieh Emami acknowledges the support from grant numbers 21-atp21-0077, NSF AST-1816420, and HST-GO-16173.001-A as well as the Institute for Theory and Computation at the Center for Astrophysics. Support for this work was provided by the NSF through grants AST-1952099, AST-1935980, AST-1828513, and by the Gordon and Betty Moore Foundation through grant GBMF-10423. This work has been supported in part by the Black Hole Initiative at Harvard University, which is funded by grants from the John Templeton Foundation and the Gordon and Betty Moore Foundation to Harvard University. This work received guidance from AstroAI at the Center for Astrophysics - Harvard \& Smithsonian. 

\section*{Data Availability}
The primary raytraced images presented in this work is available on request to CMF at \href{mailto:christian.fromm@uni-wuerzburg.de}{christian.fromm@uni-wuerzburg.de}. The code used for generating these images is available on GitHub at \url{https://github.com/aryamohan23/EHT-GANs}.


\bibliographystyle{mnras}
\bibliography{paper}





\begin{table*}
\label{tab:mytable}
\centering
\caption{Architecture used for calculation of TSTR and TRTR}
\begin{tabular}{|c|c|c|c|c|c|c|}
            \hline
            \multicolumn{2}{|c|}{\textbf{Operation Layer}} & \textbf{\begin{tabular}[c]{@{}c@{}}Number \\ of Filters\end{tabular}} & \textbf{\begin{tabular}[c]{@{}c@{}}Size of \\ Each Filter\end{tabular}} & \textbf{\begin{tabular}[c]{@{}c@{}}Stride\\ Value\end{tabular}} & \textbf{\begin{tabular}[c]{@{}c@{}}Padding\\ Value\end{tabular}} & \textbf{\begin{tabular}[c]{@{}c@{}}Size of \\ Output Image\end{tabular}} \\ 
            \hline \hline
            \multicolumn{2}{|c|}{\textbf{Input Image}} & - & - & - & - & $160 \times 160 \times 2$ \\ 
            \hline
            \textbf{Convolution Layer} & Convolution & 10 & $3 \times 3$ & $1 \times 1$ & $1 \times 1$ & $160 \times 160 \times 10$ \\ 
            \cline{2-7}
            & Leaky Relu & - & - & - & - & $160 \times 160 \times 10$ \\ 
            \hline
            \textbf{Pooling Layer} & Max Pooling & - & - & - & - & $80 \times 80 \times 10$ \\ 
            \hline
            \textbf{Convolution Layer} & Convolution & 20 & $3 \times 3$ & $1 \times 1$ & $1 \times 1$ & $80 \times 80 \times 20$ \\ 
            \cline{2-7}
            & Leaky Relu & - & - & - & - & $80 \times 80 \times 20$ \\ 
            \hline
            \textbf{Pooling Layer} & Max Pooling & - & - & - & - & $40 \times 40 \times 20$ \\ 
            \hline
            \textbf{Convolution Layer} & Convolution & 50 & $3 \times 3$ & $1 \times 1$ & $1 \times 1$ & $40 \times 40 \times 50$ \\ 
            \cline{2-7}
            & Leaky Relu & - & - & - & - & $40 \times 40 \times 50$ \\ 
            \hline
            \textbf{Pooling Layer} & Max Pooling & - & $3 \times 3$ & $1 \times 1$ & $1 \times 1$ & $20 \times 20 \times 50$ \\ 
            \hline
            \textbf{Convolution Layer} & Convolution & 50 & $3 \times 3$ & $1 \times 1$ & $1 \times 1$ & $20 \times 20 \times 50$ \\ 
            \cline{2-7}
            \textbf{(two times)} & Leaky Relu & - & - & - & - & $20 \times 20 \times 50$ \\ 
            \hline
            \textbf{Pooling Layer} & Max Pooling & - & $3 \times 3$ & $1 \times 1$ & $1 \times 1$ & $10 \times 10 \times 50$ \\ 
            \hline
            \textbf{Convolution Layer} & Convolution & 50 & $3 \times 3$ & $1 \times 1$ & $1 \times 1$ & $10 \times 10 \times 50$ \\ 
            \cline{2-7}
            \textbf{(two times)} & Leaky Relu & - & - & - & - & $10 \times 10 \times 50$ \\ 
            \hline
            \textbf{Pooling Layer} & Avg Pooling & - & $10 \times 10$ & $1 \times 1$ & $1 \times 1$ & $1 \times 1 \times 50$ \\ 
            \hline
            \textbf{FC Layer} & Fully Connected & - & - & - & - & 1 \\ 
            \hline
        \end{tabular}
\end{table*}

\section*{Appendix}
The architecture seen in Table \ref{tab:mytable} used by \citep{tao} was used to evaluate our model as  it has been shown to be effective in relating spatial features in black hole images to parameters such as $a_*$ and $R_{high}$. 



\bsp	
\label{lastpage}
\end{document}